%% file: main.tex
\documentclass[aps,prd,twocolumn,nofootinbib,superscriptaddress,10pt,floatfix]{revtex4-2}
\usepackage{amsmath,amssymb,mathtools,bm}
\usepackage{graphicx,booktabs,array}
\usepackage[dvipsnames]{xcolor}
\usepackage[colorlinks=true,linkcolor=MidnightBlue,citecolor=BrickRed,urlcolor=RoyalBlue]{hyperref}
\usepackage{microtype}
\graphicspath{{figures/}}
\newcommand{\Mp}{M_{\rm P}}
\newcommand{\Pz}{\mathcal P_{\zeta}}

\newcommand{\dd}{\mathrm d}
\newcommand{\Fig}[1]{\hyperref[#1]{Fig.~\ref*{#1}}}
\newcommand{\Sec}[1]{\hyperref[#1]{Sec.~\ref*{#1}}}
\newcommand{\Eq}[1]{\hyperref[#1]{Eq.~(\ref*{#1})}}
\newcommand{\App}[1]{\hyperref[#1]{App.~\ref*{#1}}}
\newcommand{\Tab}[1]{\hyperref[#1]{Tab.~\ref*{#1}}}
\begin{document}

\title{Primordial black holes and scalar-induced gravitational waves in the supersymmetric $331$ model from $SU(6)$}

\author{Imtiaz Khan}
\email{ikhanphys1993@gmail.com}
\affiliation{Department of Physics, Zhejiang Normal University, Jinhua, Zhejiang 321004, China}
\affiliation{Research Center of Astrophysics and Cosmology, Khazar University, Baku, AZ1096, 41 Mehseti Street, Azerbaijan}

\author{Niamat Ullah}
\email{Niamat.ullah@buitms.edu.pk}
\affiliation{Department of Physics, Balochistan University of Information Technology, Engineering and Management Sciences (BUITEMS), Quetta, Balochistan, Pakistan.}

\author{Ali Muhammad}
\email{alimuhammad@phys.qau.edu.pk}
\affiliation{CAS Key Laboratory of Theoretical Physics, Institute of Theoretical Physics, Chinese Academy of Sciences, Beijing 100190, China}
\affiliation{School of Physical Sciences, University of Chinese Academy of Sciences, No. 19A Yuquan Road, Beijing 100049, China}

\author{Mussawir Khan} 
\email{mussawirkhan@ihep.ac.cn}
\affiliation{State Key Laboratory of Particle Astrophysics, Institute of High Energy Physics, Chinese Academy of Sciences, Beijing 100049, China}
\affiliation{University of Chinese Academy of Sciences, Beijing 100049, China}

\author{G. Mustafa}
\email{gmustafa3828@gmail.com}
\affiliation{Department of Physics, Zhejiang Normal University, Jinhua, Zhejiang 321004, China}

\author{Farruh~Atamurotov}
\email{atamurotov@yahoo.com}
\affiliation{Kimyo International University in Tashkent, Shota Rustaveli str. 156, Tashkent 100121, Uzbekistan}

\author{Ahmadjon~Abdujabbarov}
\email{ahmadjonab@gmail.com}
\affiliation{School of Physics, Harbin Institute of Technology, Harbin 150001, People's Republic of China}

\begin{abstract}

We study primordial black holes and scalar-induced gravitational waves in the supersymmetric $331$ model based on $SU(6)$. The inflaton continues to be the $D$-flat radial combination of the $331$ symmetry-breaking fields $\Phi$ and $\bar\Phi$, fields which are already accountable for the CMB branch. A gauge-invariant contribution that is localised in the K\"ahler frame function causes the force in the Einstein frame to decrease over a finite range of the field and results in a temporary nonattractor phase. We calculate the supergravity potential and the radial metric, evolve the background and the Mukhanov-Sasaki modes, and assess the neutral covariant Hessian along the local region. The CMB trajectory is still in agreement with the present observational constraints while the curvature spectrum attains a value of $\mathcal P_{\zeta}\simeq5.38\times10^{-3}$ when $k\simeq1.88\times10^{13}\,{\rm Mpc}^{-1}$. The enhanced spectrum leads to a primordial black hole distribution with a mass comparable to that of an asteroid and to a maximum in the scalar-induced gravitational wave signal at a frequency of $3.17\times10^{-2}\,{\rm Hz}$ with $\Omega_{\rm GW,0}h^2\simeq9.41\times10^{-11}$. The local geometry also results in a clear hierarchy of sensitivities. The gravitational wave spectrum changes smoothly as one moves along nearby trajectories whereas the Gaussian tail of the primordial black hole distribution changes much more rapidly. The positive eigenvalues of the neutral transverse Hessian remain throughout the reference trajectory. A Gaussian critical collapse calculation yields a reference integrated primordial black hole fraction of near $9.08\times10^{-3}$, while recent nonlinear studies suggest significant corrections to the statistics of rare collapses at the scalar power achieved here.

\end{abstract}

\maketitle

\section{Introduction}

Primordial black holes and scalar-induced gravitational waves allow us to study primordial fluctuations on scales that lie well beyond the direct reach of the CMB, together with later relic production mechanisms that are seeded by primordial inhomogeneities~\cite{Pirzada:2026npl}. A temporary deviation from the slow roll attractor can reduce the inflaton velocity, enhance the curvature perturbations, and thus lead to the formation of primordial black holes after horizon reentry as well as to the generation of a stochastic tensor background at second order~\cite{GarciaBellidoRuizMorales2017,MotohashiHu2017,ByrnesColePatil2019,AnandaClarksonWands2007,BaumannSteinhardtTakahashiIchiki2007,KohriTerada2018,Domenech2021}. The two observable quantities react in different ways to the enhanced spectrum: scalar-induced gravitational waves mainly depend on the scalar two point function, while the abundance of primordial black holes examines the rare probability tail along with the collapse profile and the primordial statistics~\cite{YoungByrnesSasaki2014,Musco2019,YoungMusso2020,DeLucaEtAl2019,BaguiEtAl2025}.

Several mechanisms have already been realized in the context of supersymmetry. In the case of hybrid and no scale supergravity, a local Kähler structure is able to produce a brief nonattractor phase and an enhancement of small scale power~\cite{SpanosStamou2021,WuGongLi2021}. Curved field space motion can lead to amplification in nonminimal supergravity~\cite{KawaiKim2023}. The dynamics of a supersymmetric waterfall can link the amplified scales to particle masses and to the reheating process~\cite{AfzalGhoshalKing2025,Khan:2023snv}. GUT Higgs models offer a more direct structural analogy since a field responsible for symmetry breaking is also involved in inflation~\cite{IjazRehman2025,IjazMehmoodRehman2025,Ijaz:2026ear,MajiMoursyShafi2025}. Current results show that the production of PBHs is viable in the context of supersymmetric inflation and at the same time identify a more precise model building issue. A localized amplification becomes more predictive when it acts on an inflationary direction that is already constrained by an independent particle physics sector rather than on a field that has been introduced solely for the purpose of producing the small scale feature.

A viable broad plateau is obtained by the parent nonminimal trajectory, it remaining too smooth to produce significant short-scale amplification, in agreement with the general role of controlled geometric deformations in the construction of inflationary models~\cite{Pirzada:2026emt}. The supersymmetric $331$ model based on $SU(6)$ offers such a scenario~\cite{KhanLi2026,Khan:2026zuj}. Common multiplets link the inverse seesaw, gauge mediation, unified breaking, and inflationary sectors. The pair $\Phi$ and $\bar\Phi$ is responsible for breaking the intermediate $331$ symmetry and for forming the inflationary radial mode. The coupling which governs $331$ breaking also determines the CMB normalisation and the radial mass in the broken vacuum. A local K\"ahler contribution, constructed from the gauge invariant bilinear $B=\bar\Phi\Phi/M_{\rm P}^2$, alters the radial force only within a narrow range and restores the parent geometry away from the feature.

The extent of this effect is determined by the effective radial force $G^{-1}V_{,y}/V$. It is therefore possible for a small local variation in the potential to cause a much larger change in the velocity as the logarithmic slope decreases. The time that particles spend in the shallow region determines the growth of the superhorizon curvature. The scalar source that results enters the GW convolution in a smooth manner, while the probability of PBH formation amplifies small changes via the rare collapse tail. Recent lattice calculations show that nonlinear corrections of order several percent to tens of percent appear when the scalar spectrum reaches the range $\mathcal O(10^{-3})$ to $\mathcal O(10^{-2})$~\cite{Caravano2025,Caravano2025II}. The perturbative two point spectrum thus serves as a well-defined baseline for the GW calculation and as a Gaussian reference for the PBH abundance.

We directly derive the noncanonical background and perturbation modes and look at the neutral supergravity Hessian over the entire localized interval. Instead of reducing the comparison to individual markers, the continuous CMB trajectories reveal the pivot dependence. The numerical scalar spectrum is then used in the radiation era GW convolution as well as in the critical collapse PBH calculation. Section~\ref{sec:model} deals with the particle sector and Kähler geometry. Section~\ref{sec:bg} describes the background and perturbation dynamics. Section~\ref{sec:stability} investigates the neutral stability. Section~\ref{sec:scalar} gives the scalar spectrum and the CMB trajectories. Sections~\ref{sec:gw} and \ref{sec:pbh} cover the two small-scale observables. Section~\ref{sec:comparison} compares the resulting mechanism with other nearby supersymmetric models.

\section{Particle physics structure and inflation in the supersymmetric $331$ model from $SU(6)$}\label{sec:model}

\subsection{$SU(6)$ embedding, matter assignment, and selection rules}

The particle physics community is following the supersymmetric $331$ model based on $SU(6)$~\cite{KhanLi2026}. The fundamental normalization of the Abelian generator of the intermediate gauge group is

\begin{equation}
\begin{aligned}
T_X&=\frac{1}{2\sqrt3}\,\mathrm{diag}(-1,-1,-1,1,1,1),\\
\mathrm{Tr}\,T_X^2&=\frac12 .
\end{aligned}
\label{eq:TX}
\end{equation}

With $q_6\equiv1/(2\sqrt3)$, the fundamental and antisymmetric representations branch according to

\begin{align}
\mathbf 6&=(\mathbf3,\mathbf1,-q_6)\oplus(\mathbf1,\mathbf3,+q_6),\label{eq:branch6new}\\
\mathbf{15}&=(\overline{\mathbf3},\mathbf1,-2q_6)\oplus(\mathbf1,\overline{\mathbf3},+2q_6)\oplus(\mathbf3,\mathbf3,0).\label{eq:branch15new}
\end{align}

The electric charge operator is

\begin{equation}
Q=T_{3L}+\frac{1}{\sqrt3}T_{8L}+\frac{2}{\sqrt3}X,
\label{eq:Qembed}
\end{equation}

The right-handed neutrino sector of the $331$ models is realized by the embedding~\cite{PisanoPleitez1992,Frampton1992,Montero1993,FootLongTran1994,Long1996,Deppisch2016,LiPeiXuZhang2020}. Each chiral family is then placed in

\begin{equation}
\mathcal F_i=\mathbf{15}_i\oplus\overline{\mathbf6}_i\oplus\overline{\mathbf6}'_i,
\qquad i=1,2,3,
\label{eq:familynew}
\end{equation}

By using the values $A(\mathbf{6}) = 1$ and $A(\mathbf{15}) = 2$, the cubic $SU(6)$ anomaly cancels one family at a time.

\begin{equation}
A(\mathcal F_i)=2-1-1=0.
\label{eq:anomnew}
\end{equation}

The Higgs sector and the messenger sector are vectorlike with respect to $SU(6)$; the fields relevant to the operators are listed in \Tab{tab:fieldsnew}, and their assignment already leads to the inverse seesaw, gauge mediation, two-step unified breaking, and the inflationary direction prior to the PBH sector being introduced~\cite{KhanLi2026}.

The superpotential carries a $Z_4^R$ charge of two; the assignments that are relevant for the matter, Higgs, and 331 sectors are

\begin{align}
R(15_i,\bar6_i,\bar6'_i)&=1, & R(N_s)&=3,\nonumber\\
R(H_{15},\bar H_6)&=0, & R(\bar H_{15},H_6)&=2,\nonumber\\
R(\Phi,\bar\Phi,S)&=2.&&
\label{eq:Rchargesnew}
\end{align}

The fields $\bar6'_i$ and $N_{si}$ are odd with respect to $Z_2^\nu$ and all the other fields listed in \Tab{tab:fieldsnew} are even. The adjoint shaping symmetry gives the $Z_3^G$ charge one to the fields $A$, $B$, and $Z$, whereas the light matter and Higgs multiplets have no charge. These selection rules are treated as effective discrete symmetries of the unified theory~\cite{LeeR2011,ChenLiuTeng2021} and yield the renormalizable terms.

\begin{align}
W={}&\frac14y^u_{ij}15_i15_jH_{15}
+\sqrt2y^d_{ij}15_i\bar6_j\bar H_6\nonumber\\
&+\frac14y^\nu_{ij}\bar6'_i\bar6'_jH_{15}
+Y_{ij}\bar6'_i\Phi N_{s j}\nonumber\\
&+\kappa S\!\left(\bar\Phi\Phi-\frac{v_0^2}{2}\right)
+W_H+W_G+W_{\rm GM}.
\label{eq:fullWnew}
\end{align}

For example, $R(\bar6'\Phi N_s)=1+2+3=2\pmod4$ and $R(S\bar\Phi\Phi)=2+2+2=2\pmod4$, even though both operators are even under $Z_2^\nu$. It is therefore the combination of gauge invariance with the discrete charges which selects the inverse seesaw and $331$ breaking operators used below. The fact that there is an antisymmetric contraction in $\bar6'\bar6'H_{15}$ means that $y^{\nu T}=-y^\nu$ and leads to the leading rank two inverse seesaw texture of the parent model~\cite{KhanLi2026}.

The other parts link the inflationary sector to the particle physics model, and the Higgs mixing, adjoint breaking, and gauge mediation terms can be expressed as

\begin{align}
W_H={}&\mu_{15}\bar H_{15}H_{15}+\mu_6\bar H_6H_6\nonumber\\
&+\lambda_HH_{15}\bar H_6\bar\Phi
+\bar\lambda_H\bar H_{15}H_6\Phi,\label{eq:WHnew}\\
W_G={}&\lambda_ZZ\left(\mathrm{Tr}A^2-\frac{v_A^2}{2}\right)\nonumber\\
&+\lambda_B\mathrm{Tr}\!\left[B\left(A^2-\frac{\mathrm{Tr}A^2}{6}\mathbf1_6\right)\right],\label{eq:WGnew}\\
W_{\rm GM}={}&\sum_{A=1}^{2}\lambda_A X\Psi_A\bar\Psi_A,\nonumber\\
&\langle X\rangle=M_{\rm GM}+\theta^2F_X .
\label{eq:WGMnew}
\end{align}
\begin{table*}[t]
\centering
\caption{The superfields of the supersymmetric 331 model associated with the particle physics and inflationary sectors have an R value corresponding to the $Z_4^R$ charge and a $P_\nu$ value corresponding to the $Z_2^\nu$ parity; for the vectorlike pairs, the conjugate $G_{331}$ components are understood.}
\label{tab:fieldsnew}
\scriptsize
\resizebox{\textwidth}{!}{%
\begin{tabular}{@{}llllll@{}}
\toprule
superfield & $SU(6)$ & representative $G_{331}$ content & $R$ & $P_\nu$ & role \\
\midrule
$15_i$ & $\mathbf{15}$ & $(\mathbf3,\mathbf3,0)\oplus(\overline{\mathbf3},\mathbf1,-1/\sqrt3)\oplus(\mathbf1,\overline{\mathbf3},+1/\sqrt3)$ & $1$ & $+$ & quarks, charged leptons, $\nu^c$ components \\
$\bar6_i$ & $\overline{\mathbf6}$ & $(\overline{\mathbf3},\mathbf1,+q_6)\oplus(\mathbf1,\overline{\mathbf3},-q_6)$ & $1$ & $+$ & ordinary antifundamental matter \\
$\bar6'_i$ & $\overline{\mathbf6}$ & $(\overline{\mathbf3},\mathbf1,+q_6)\oplus(\mathbf1,\overline{\mathbf3},-q_6)$ & $1$ & $-$ & inverse seesaw lepton multiplet \\
$N_{s i}$ & $\mathbf1$ & $(\mathbf1,\mathbf1,0)$ & $3$ & $-$ & inverse seesaw singlet \\
$H_{15}+\bar H_{15}$ & $\mathbf{15}+\overline{\mathbf{15}}$ & electroweak triplet/antitriplet fragments & $0,2$ & $+$ & up type and neutrino operators \\
$H_6+\bar H_6$ & $\mathbf6+\overline{\mathbf6}$ & electroweak triplet/antitriplet fragments & $2,0$ & $+$ & down type operators \\
$\Phi+\bar\Phi$ & $\mathbf6+\overline{\mathbf6}$ & $(\mathbf1,\mathbf3,+q_6)\oplus(\mathbf1,\overline{\mathbf3},-q_6)$ & $2,2$ & $+$ & $331$ breaking and inflation \\
$S$ & $\mathbf1$ & $(\mathbf1,\mathbf1,0)$ & $2$ & $+$ & $F$ term breaking/inflation singlet \\
$A,B$ & $\mathbf{35}$ & adjoint fragments & $0,2$ & $+$ & $SU(6)\to G_{331}$ vacuum \\
$Z$ & $\mathbf1$ & $(\mathbf1,\mathbf1,0)$ & $2$ & $+$ & adjoint constraint \\
$X$ & $\mathbf1$ & spurion & $2$ & $+$ & gauge mediation spurion \\
$\Psi_A+\bar\Psi_A$ & $\mathbf6+\overline{\mathbf6}$ & complete messenger multiplet & $0,0$ & $+$ & gauge mediation, $A=1,2$ \\
\bottomrule
\end{tabular}%
}
\end{table*}

The adjoint vacuum, when it aligns with $T_X$, leaves the $G_{331}$ symmetry unbroken, whereas the lower scale pair $\Phi + \bar\Phi$ later breaks the $331$ symmetry. This same pair is found in \Eq{eq:fullWnew}, in the inverse seesaw operator, and in the inflationary $D$ flat direction. It is therefore the shared field content that connects the CMB normalization to the physical radial state of the $331$ breaking sector rather than bringing in an unrelated inflation parameter.

\subsection{$331$ breaking inflationary trajectory}

In the supersymmetric $331$ model based on $SU(6)$, the fields $\Phi$ and $\bar\Phi$ form a $\mathbf 6+\overline{\mathbf 6}$ pair and it is their neutral components that cause the $331$ symmetry to be broken along a $D$ flat direction. The singlet $S$ is responsible for this breaking via the inflationary superpotential~\cite{KhanLi2026}.

\begin{equation}
W_{\rm inf}=\kappa S\left(\bar\Phi\Phi-\frac{v_0^2}{2}\right),
\label{eq:W}
\end{equation}

The real $D$ flat trajectory is parameterized by

\begin{equation}
\Phi=\bar\Phi=\frac{\varphi}{2},\qquad S=0,
\label{eq:dflat}
\end{equation}

With $\phi$ being real, the original inflationary Kähler potential is

\begin{multline}
K_0=-3\Mp^2\ln \Omega_0,\\
\Omega_0=1-\frac{|\Phi|^2+|\bar\Phi|^2+|S|^2}{3\Mp^2}
+\frac{\chi(\bar\Phi\Phi+{\rm h.c.})}{2\Mp^2}
+\frac{\zeta_S|S|^4}{3\Mp^4}.
\label{eq:K0}
\end{multline}

Defining $y\equiv\varphi/\Mp$, $\nu\equiv v_0/\Mp$, and $\xi\equiv\chi/4-1/6$, the real frame function on \Eq{eq:dflat} is

\begin{equation}
F_0(y)=1+\xi y^2.
\end{equation}

The original model takes $\xi=100$ as its benchmark for a large $\xi$ plateau~\cite{KhanLi2026}. The nonminimal version is part of the broader class of supergravity inflation discussed in~\cite{KalloshLinde2010,BezrukovShaposhnikov2008,Moursy2021}.

The supergravity $F$ term potential is
\begin{equation}
\begin{aligned}
V_F&=e^{K/\Mp^2}\left(K^{I\bar J}D_IW D_{\bar J}\bar W-\frac{3|W|^2}{\Mp^2}\right),\\
D_IW&=\partial_IW+\frac{K_I}{\Mp^2}W.
\end{aligned}
\label{eq:VF}
\end{equation}

Along \Eq{eq:dflat}, $W=0$ and $D_\Phi W=D_{\bar\Phi}W=0$, while
\begin{equation}
D_SW=\kappa\left(\bar\Phi\Phi-\frac{v_0^2}{2}\right).
\end{equation}

On the trajectory with S equal to 0 and B equal to B dagger, the phase-stabilizing term, which is introduced below, disappears together with its first radial pullback derivatives, and it follows that K to the S bar S equals F. Equations \Eq{eq:VF} and \Eq{eq:dflat} yield therefore

\begin{equation}
V_E(y)=\frac{\kappa^2\Mp^4(y^2-2\nu^2)^2}{16F(y)^2}.
\label{eq:VE}
\end{equation}

When F tends to $F_0$, the parent potential is obtained.

The same $\kappa$ also enters the broken phase radial mass relation in~\cite{KhanLi2026},

\begin{equation}
m_\rho=\kappa v_{331}.
\label{eq:mrho}
\end{equation}

The physical mass in the symmetry breaking sector thus becomes associated with the normalization of the CMB for a given $331$ scale; in the case of $v_{331}=30\,{\rm TeV}$, the benchmark normalization obtained below, $\kappa\simeq4.07\times10^{-3}$, implies that $m_\rho\simeq122\,{\rm GeV}$, a value which is close to that found in the original analysis for its $N_*=55$ benchmark.

\subsection{Covariant localized K\"ahler operator}

We define the gauge invariant dimensionless bilinear
\begin{equation}
B\equiv \frac{\bar\Phi\Phi}{\Mp^2}
\end{equation}
and write the localized frame function as
\begin{multline}
\Omega=\Omega_0+\frac{A}{2}\left[f(B)+f(B)^\dagger\right]
+\frac{c_\theta}{\sigma_u^2}\frac{|S|^2}{\Mp^2}(B-B^\dagger)^2,\\
f(B)=\operatorname{sech}^2\left(\frac{2B-u_0}{\sigma_u}\right).
\label{eq:OmegaFeature}
\end{multline}

The initial local contribution consists of the holomorphic and antiholomorphic parts and does not give rise to any additional mixed derivative involving a nonlinear function of $B+B^\dagger$. The final term disappears when $S=0$ on the real trajectory as well as with its first radial pullback derivatives. The radial equations \Eq{eq:F} and \Eq{eq:G} stay the same, but deviations from the trajectory affect the neutral transverse sector via the metric of the $S$ field. The stability calculation mentioned below uses $c_\theta=-5\times10^{-4}$. Along \Eq{eq:dflat}, \Eq{eq:OmegaFeature} reduces to
\begin{equation}
F(y)=1+\xi y^2+A\operatorname{sech}^2\left(\frac{y^2/2-u_0}{\sigma_u}\right).
\label{eq:F}
\end{equation}

The radial kinetic term is
\begin{equation}
\mathcal L_{\rm kin}=\frac{\Mp^2}{2}G(y)(\partial y)^2,
\qquad
G(y)=\frac{1}{F}+\frac{3}{2}\left(\frac{F_{,y}}{F}\right)^2.
\label{eq:G}
\end{equation}

The limit in which $A$ tends to zero yields the original inflationary trajectory of the supersymmetric $331$ model from $SU(6)$ exactly; the localized structure thus alters only a finite interval of the K\"ahler geometry and leaves the asymptotic nonminimal plateau unmodified.

For fixed $(u_0,\sigma_u)$, a stationary inflection reference pair $(y_c,A_c)$ obeys

\begin{equation}
V_{,y}(y_c;A_c)=0,\qquad V_{,yy}(y_c;A_c)=0,
\label{eq:Ac}
\end{equation}
and parameterize the traversing solutions by
\begin{equation}
\Delta_K\equiv \frac{A}{A_c}-1.
\label{eq:DeltaK}
\end{equation}

When $u_0$ is 0.18 and $\sigma_u$ is 0.002, the numerical solution is
\begin{equation}
y_c=0.597801,\qquad A_c=1.46825\times10^{-2}.
\label{eq:Acnum}
\end{equation}

If we write the canonical radial coordinate in the form $\dd\chi=\Mp\sqrt{G}\,\dd y$, then the local critical structure becomes apparent. The expansion of the dimensionless potential $U=V_E/(\kappa^2\Mp^4)$ about the point $(\chi_c,A_c)$ is as follows:

\begin{equation}
U_{,\chi}=C_A\delta A+\frac{C_3}{2}\delta\chi^2
+\mathcal O(\delta A\delta\chi,\delta\chi^3),
\label{eq:normalform}
\end{equation}
where $\delta A$ is equal to $A - A_c$ and $\delta\chi$ is equal to $\chi - \chi_c$. The coefficients satisfy $C_A = \left. \frac{\partial U}{\partial A} \right|_c$ and $C_3 = \left. \frac{\partial^3 U}{\partial \chi^3} \right|_c$. On the branch which contains two nearby stationary points,

\begin{equation}
\delta\chi_{\pm}=\pm\left(-\frac{2C_A\delta A}{C_3}\right)^{1/2}
+\mathcal O(\delta A).
\label{eq:saddlescale}
\end{equation}

The stationary separation therefore scales as $|\delta A|^{1/2}$ and integration of the local force gives a barrier scale proportional to $|\delta A|^{3/2}$. The background kinetic energy distinguishes traversal from trapping. Near the traversing boundary, a small amplitude change can consequently generate a much larger change in residence time before the trajectory recovers the parent slope.

\begin{figure*}[t]
\centering
\includegraphics[width=0.92\textwidth]{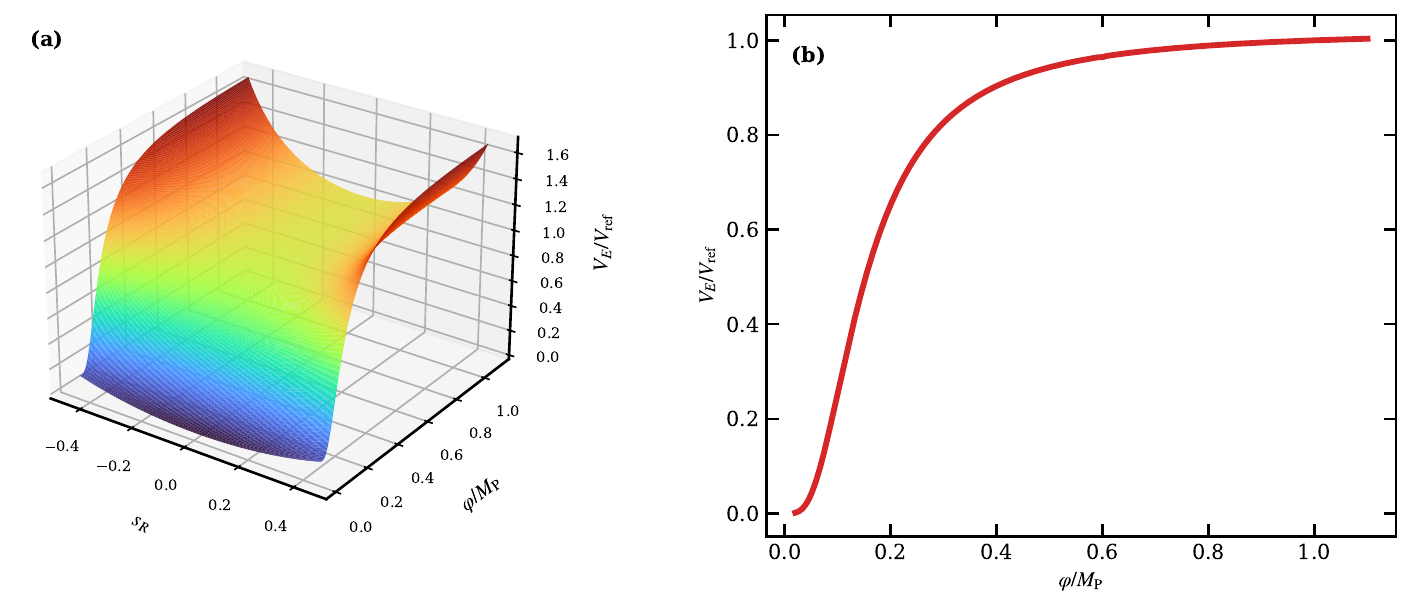}
\caption{Normalized neutral Einstein frame potential for the reference K\"ahler branch. The left panel shows the real $(s_R,\varphi)$ surface with $\Phi=\bar\Phi=\varphi/2$ and $S=s_R\Mp/\sqrt2$. The right panel shows the inflationary slice at $s_R=0$. The smooth valley connects the symmetry-breaking direction to the nonminimal plateau.}
\label{fig:modelpotential}
\end{figure*}

\Fig{fig:modelpotential} shows the neutral Einstein frame potential before the local feature is examined in detail. Parameterizing the real singlet direction by $S=s_R\Mp/\sqrt2$ and retaining $\Phi=\bar\Phi=\varphi/2$ gives the displayed surface from the complete neutral supergravity potential $V_F+V_D$ in \Eq{eq:VF} and \Eq{eq:OmegaFeature}. The right panel gives its $s_R=0$ slice. The inflationary valley lies along the singlet origin, while the radial direction approaches the nonminimal plateau.

\begin{figure}[t]
\centering
\includegraphics[width=0.98\columnwidth]{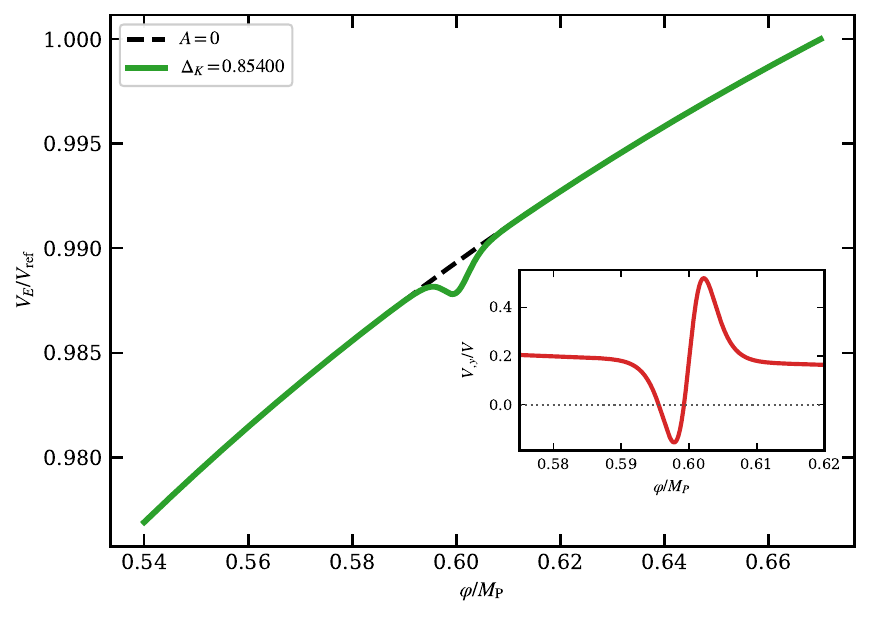}
\caption{The Einstein frame potential near the local Kähler structure is shown; the dashed curve represents the parent A=0 trajectory and the green curve corresponds to the PBH benchmark. The inset displays the logarithmic slope $V_{,y}/V$; the temporary decrease in slope delays the radial motion and results in a nonattractor interval.}
\label{fig:geometry}
\end{figure}

The $\Delta_K$ value for the PBH reference branch is 0.85400; the local geometry is given in \Fig{fig:geometry} and the potential undergoes only a slight change. The key dynamical quantity is the temporary decrease in the logarithmic slope as shown in the inset.

The figure referred to as \Fig{fig:geometry} shows the dynamical effect of the local K\"ahler contribution while maintaining the original nonminimal plateau. Throughout most of the field range, the benchmark conforms to the parent $A=0$ potential. In a narrow interval, there is only a small alteration of $V_E$ but a much greater decrease in $G^{-1}V_{,y}/V$. As a result, the inflaton loses velocity for a finite duration. The recovery of the parent slope then brings the trajectory back to the normal slow roll branch.

\section{Background and perturbation evolution}\label{sec:bg}
Writing $N\equiv\ln a$ and $p\equiv\dd y/\dd N$, the first Hubble flow parameter is

\begin{equation}
\epsilon_H=\frac12 Gp^2,
\label{eq:eps}
\end{equation}
and the Friedmann equation becomes
\begin{equation}
H^2=\frac{V}{\Mp^2(3-\epsilon_H)}.
\label{eq:H}
\end{equation}

The exact homogeneous equation is
\begin{equation}
p'+\Gamma^y_{yy}p^2+(3-\epsilon_H)\left[p+\frac1G\frac{V_{,y}}V\right]=0,\qquad
\Gamma^y_{yy}=\frac12\frac{G_{,y}}G.
\label{eq:bg}
\end{equation}

Inflation stops when $\epsilon_H$ reaches 1. It is useful to define
\begin{equation}
\epsilon_2\equiv\frac{\dd\ln\epsilon_H}{\dd N},
\qquad
\eta_H\equiv\epsilon_H-\frac12\epsilon_2.
\end{equation}

\begin{figure}[t]
\centering
\includegraphics[width=0.98\columnwidth]{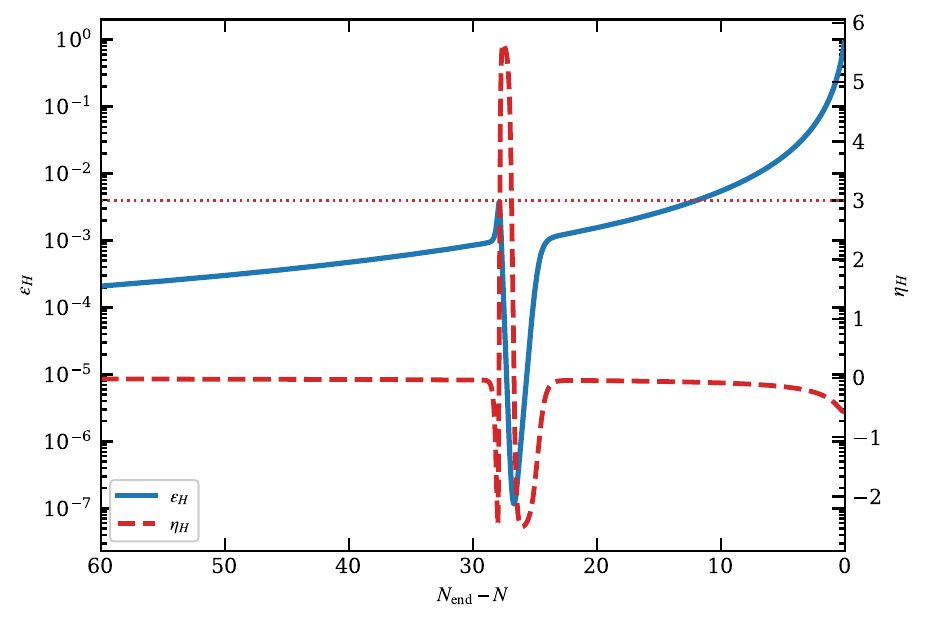}
\caption{Hubble flow parameters for the PBH benchmark. The localized K\"ahler feature produces a finite nonattractor interval around $N_{\rm end}-N\simeq26.7$. The horizontal dotted line marks $\eta_H=3$, the sustained ultraslow roll reference value. The finite feature overshoots the reference during the transition and then returns to the attractor.}
\label{fig:bg}
\end{figure}

For the branch involving near-critical traversal, $\epsilon_H$ decreases rapidly, and $\eta_H$ becomes positive before the trajectory reestablishes itself on the slow roll attractor. This evolution is shown in \Fig{fig:bg}.

Before reaching the localized region, $\epsilon_H$ follows the slowly varying parent attractor. Near $N_{\rm end}-N\simeq26.7$, the reduced force visible in \Fig{fig:geometry} lowers the inflaton velocity and drives the excursion in \Fig{fig:bg}. The reference $\eta_H=3$ marks sustained ultraslow roll. The finite transition overshoots the reference briefly and then returns to the ordinary attractor after the parent slope recovers. Superhorizon curvature evolution during the excursion invalidates a slow-roll estimate of the enhanced spectrum and motivates direct Mukhanov-Sasaki mode evolution.

For a single adiabatic degree of freedom, the curvature perturbation obeys
\begin{equation}
\mathcal R_{k,NN}+\left(3-\epsilon_H+\epsilon_2\right)\mathcal R_{k,N}
+\left(\frac{k}{aH}\right)^2\mathcal R_k=0.
\label{eq:MS}
\end{equation}
With $z=a\Mp\sqrt{2\epsilon_H}$, the Bunch-Davies state deep inside the horizon gives
\begin{align}
\mathcal R_k&=\frac{e^{-ik\tau}}{z\sqrt{2k}},\\
\mathcal R_{k,N}&=\left[-i\frac{k}{aH}-1-\frac{\epsilon_2}{2}\right]\mathcal R_k.
\end{align}

The mode then undergoes evolution within that local interval until its amplitude over the superhorizon becomes constant. The scalar spectrum is
\begin{equation}
\Pz(k)=\frac{k^3}{2\pi^2}|\mathcal R_k|^2.
\label{eq:Pz}
\end{equation}

The value of the CMB amplitude $\Pz(k_*)=A_s=2.10\times10^{-9}$ at $k_*=0.05\,{\rm Mpc}^{-1}$ is what determines the total coupling $\kappa$. The dimensionless background does not depend on the normalization of the overall potential, whereas the direct pivot mode fixes the final normalization.

The tensor modes are each evolved independently, for each polarization,
\begin{equation}
h_{k,NN}+(3-\epsilon_H)h_{k,N}+\left(\frac{k}{aH}\right)^2h_k=0,
\end{equation}

Using the canonical Bunch-Davies normalization, the total tensor spectrum is given by $\mathcal P_t=k^3|h_k|^2/\pi^2$. We evaluate
\begin{equation}
r=\frac{\mathcal P_t(k_*)}{\Pz(k_*)}
\end{equation}
from these numerical modes rather than from $16\epsilon_H$ alone.Similarly,
\begin{equation}
n_s-1=\left.\frac{\dd\ln\Pz}{\dd\ln k}\right|_{k_*},
\qquad
\alpha_s=\left.\frac{\dd n_s}{\dd\ln k}\right|_{k_*}
\label{eq:numericalns}
\end{equation}
are extracted by a local quadratic fit to exact scalar modes around the pivot.

\section{Neutral supergravity stability}\label{sec:stability}

The Kähler operator, when applied in a local manner, alters the field geometry in the interval that is involved in the amplification process. It thus necessitates a direct examination along the deformed trajectory, as is done in covariant treatments in which the geometry of field space determines whether or not a reduced inflationary description is valid~\cite{Pirzada:2026jml}. If the three complex neutral fields are written as $z^I=(\Phi,\bar\Phi,S)$, then the Kähler metric is obtained.

\begin{equation}
K_{I\bar J}=-3\Mp^2\left(\frac{\Omega_{I\bar J}}{\Omega}-\frac{\Omega_I\Omega_{\bar J}}{\Omega^2}\right).
\label{eq:Kmetric}
\end{equation}

The $D$ term, which is neutral, follows the broken generator normalization of the parent construction~\cite{KhanLi2026}. In six real coordinates $\phi^A$, the covariant potential Hessian satisfies
\begin{equation}
\mathcal H_{AB}=\nabla_A\nabla_BV,
\qquad
\mathcal H_{AB}u^B=m^2G_{AB}u^B.
\label{eq:Hess}
\end{equation}

Projecting away the tangent direction and the broken gauge orbit isolates the neutral transverse subspace. For $\zeta_S=1$ and $c_\theta=-5\times10^{-4}$, the real kinetic metric stays positive and the lightest transverse Hessian eigenvalue satisfies
\begin{equation}
\frac{m_{\perp,\min}^2}{H^2}>3.8
\label{eq:stabnum}
\end{equation}

\begin{figure}[t]
\centering
\includegraphics[width=0.98\columnwidth]{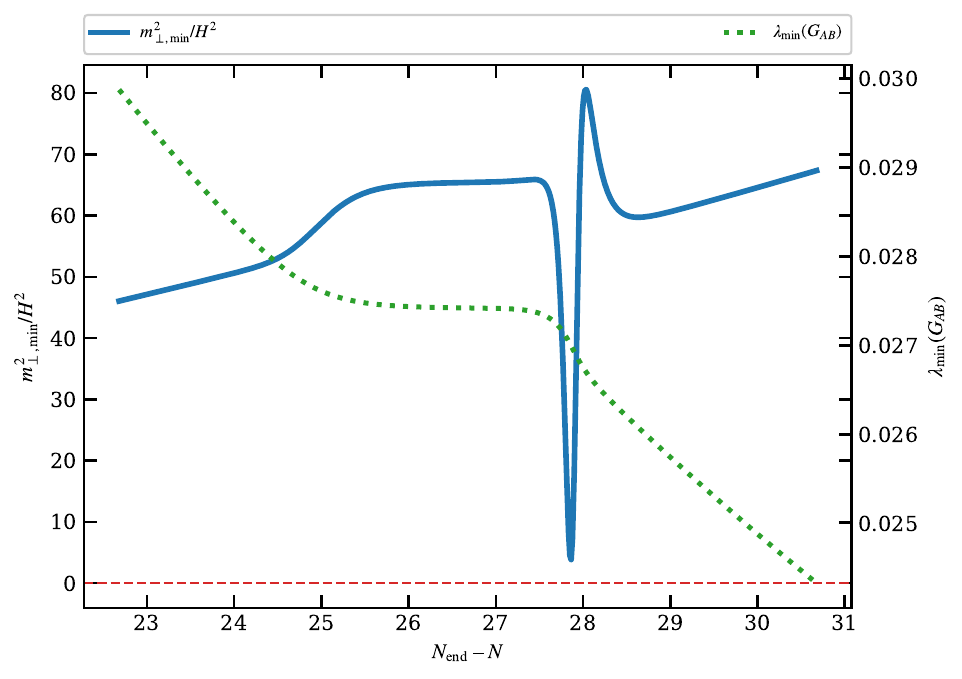}
\caption{Neutral supergravity stability through the localized interval. The blue curve shows the lightest transverse covariant Hessian eigenvalue after removing the tangent and gauge directions. The green dotted curve shows the smallest eigenvalue of the real kinetic metric. Both remain positive along the reference path.}
\label{fig:stability}
\end{figure}

In the sampled local interval, as shown in \Fig{fig:stability}, the positivity continues through the most prominent part of the feature, which supports the radial treatment within the neutral $(\Phi,\bar\Phi,S)$ sector~\cite{Pirzada:2026sle}. The result refers to the neutral covariant potential Hessian, and the full fluctuation masses for the additional unified states depend on the corresponding heavy field geometry.

\section{Scalar spectrum and CMB observables}\label{sec:scalar}

\begin{figure}[t]
\centering
\includegraphics[width=0.98\columnwidth]{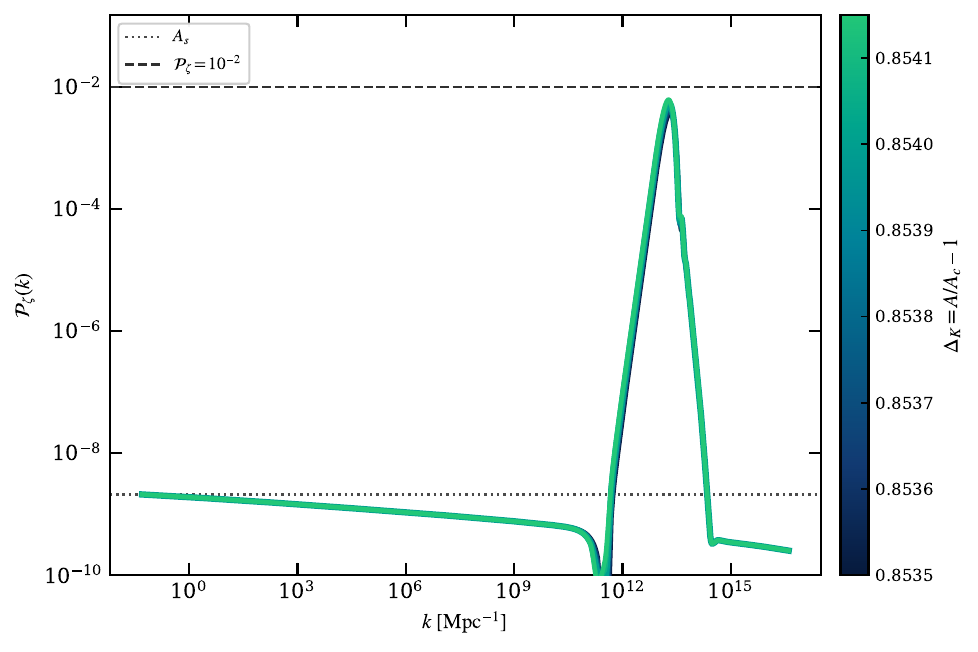}
\caption{Curvature spectra obtained by evolving the modes in \Eq{eq:MS}. The color scale encodes $\Delta_K=A/A_c-1$. Nearby K\"ahler amplitudes share the CMB branch and separate near the localized feature. Increasing $\Delta_K$ lengthens the residence time and raises the small-scale maximum while preserving its narrow structure.}
\label{fig:scalar}
\end{figure}

The feature position $u_0$, its width $\sigma_u$, and the relative amplitude $\Delta_K$ each affect different aspects of the dynamics. The position determines the time at which the encounter occurs, the width regulates the duration of the reduced force, and the amplitude controls how deep the slowdown is. Figure~\ref{fig:scalar} shows a family of nearby amplitudes at fixed $(\xi,u_0,\sigma_u)$, with their CMB branches remaining almost degenerate since the localised region appears well after pivot exit; their maxima on small scales separate quickly because a greater slowdown increases the residence time and leads to extended superhorizon growth.

For the reference branch, direct scalar and tensor mode evolution gives
\begin{align}
n_s&=0.96647, & r&=3.25\times10^{-3}, & \alpha_s&=-5.67\times10^{-4},
\label{eq:cmbnum}
\end{align}
with $\kappa=4.07\times10^{-3}$. The short-scale spectrum reaches
\begin{equation}
\Pz^{\rm peak}=5.38\times10^{-3},\qquad
k_{\rm peak}=1.88\times10^{13}\,{\rm Mpc}^{-1}.
\label{eq:peak}
\end{equation}

Close to a critical feature, the models show a similar rapid reaction of the scalar maximum to small changes in local geometry~\cite{ColeGowByrnesPatil2023}. In this case, the sequence is clear: the local force decreases, the kinetic energy drops, the growing superhorizon mode stays active for a longer time, and the recovery of the parent slope ends the amplification.

\begin{figure*}[t]
\centering
\includegraphics[width=0.98\textwidth]{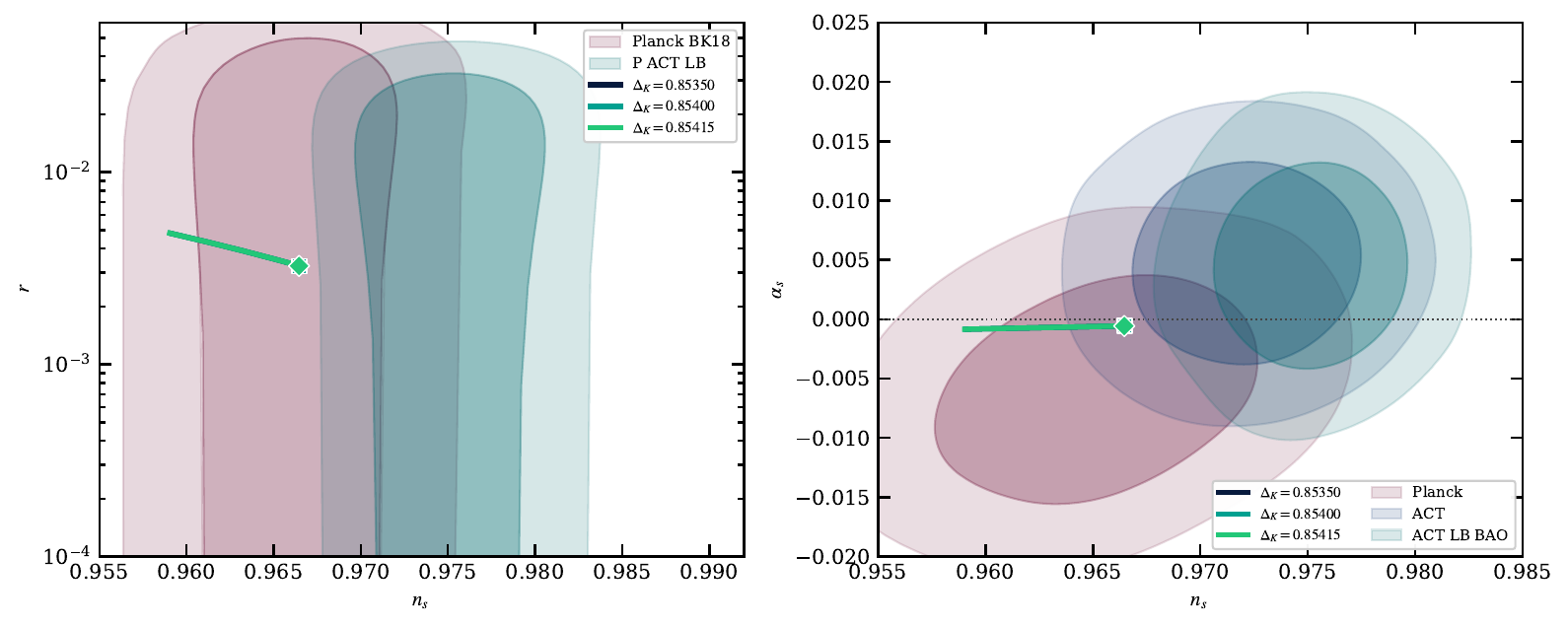}
\caption{CMB trajectories for the three K\"ahler amplitudes. The left panel shows $(n_s,r)$ with current tensor constraints~\cite{PlanckInflation2018,BICEP2021,Tristram2022,ACTDR6Power2025,ACTExtended2025}. The right panel shows $(n_s,\alpha_s)$ with Planck, ACT, and combined large scale data~\cite{PlanckInflation2018,ACTExtended2025,DESILya2026}. Each trajectory spans pivot locations from fifty to sixty e-folds before the end of inflation. The markers identify the direct mode result at the reference pivot.}
\label{fig:cmb}
\end{figure*}

\Fig{fig:cmb} compares continuous trajectories with current CMB constraints and recent nonminimal inflation analyses~\cite{PlanckInflation2018,BICEP2021,Tristram2022,ACTDR6Power2025,ACTExtended2025,DESILya2026,Pirzada:2026uak}. Each curve follows the predicted $(n_s,r,\alpha_s)$ values as the pivot ranges from fifty to sixty e-folds before the end of inflation. The paths display the pivot mapping dependence hidden by a single benchmark marker. All three feature amplitudes remain nearly degenerate over CMB scales because the localized interval is encountered later. The markers denote direct mode extraction at the reference pivot.

\section{Scalar induced gravitational waves}\label{sec:gw}

During radiation domination, the enhanced scalar perturbations source tensor modes at second order~\cite{AnandaClarksonWands2007,BaumannSteinhardtTakahashiIchiki2007,KohriTerada2018,EspinosaRaccoRiotto2018,Domenech2021}.For Gaussian scalar perturbations, the late-time radiation era result is
\begin{multline}
\Omega_{\rm GW}^{\rm RD}(k)=\frac1{12}\int_0^\infty\dd v
\int_{|1-v|}^{1+v}\dd u\\
\times\left[\frac{4v^2-(1+v^2-u^2)^2}{4uv}\right]^2
\overline{I^2(u,v)}\,\Pz(ku)\Pz(kv),
\label{eq:GWconv}
\end{multline}
where
\begin{multline}
\overline{I^2}=\frac12\left[\frac{3(u^2+v^2-3)}{4u^3v^3}\right]^2
\Bigg\{\Bigg[-4uv+(u^2+v^2-3)\\
\times\ln\left|\frac{3-(u+v)^2}{3-(u-v)^2}\right|\Bigg]^2
+\pi^2(u^2+v^2-3)^2\Theta(u+v-\sqrt3)\Bigg\}.
\label{eq:kernel}
\end{multline}

We insert the numerical spectrum from \Eq{eq:Pz} directly into \Eq{eq:GWconv}.The convolution domain is restricted to the directly evolved scalar modes. No return to $A_s$, lognormal tail, or fitted extrapolation enters outside the numerical support. Redshifting to the present gives
\begin{equation}
\Omega_{\rm GW,0}h^2=1.62\times10^{-5}
\left(\frac{g_*}{106.75}\right)
\left(\frac{g_{*s}}{106.75}\right)^{-4/3}
\Omega_{\rm GW}^{\rm RD},
\label{eq:redshift}
\end{equation}
and
\begin{equation}
f=1.55\times10^{-15}\left(\frac{k}{{\rm Mpc}^{-1}}\right){\rm Hz}.
\label{eq:kf}
\end{equation}

For the reference branch,
\begin{equation}
f_{\rm GW}^{\rm peak}=3.17\times10^{-2}\,{\rm Hz},\qquad
\Omega_{\rm GW,0}^{\rm peak}h^2=9.41\times10^{-11}.
\label{eq:gwpeak}
\end{equation}

\begin{figure}[t]
\centering
\includegraphics[width=0.98\columnwidth]{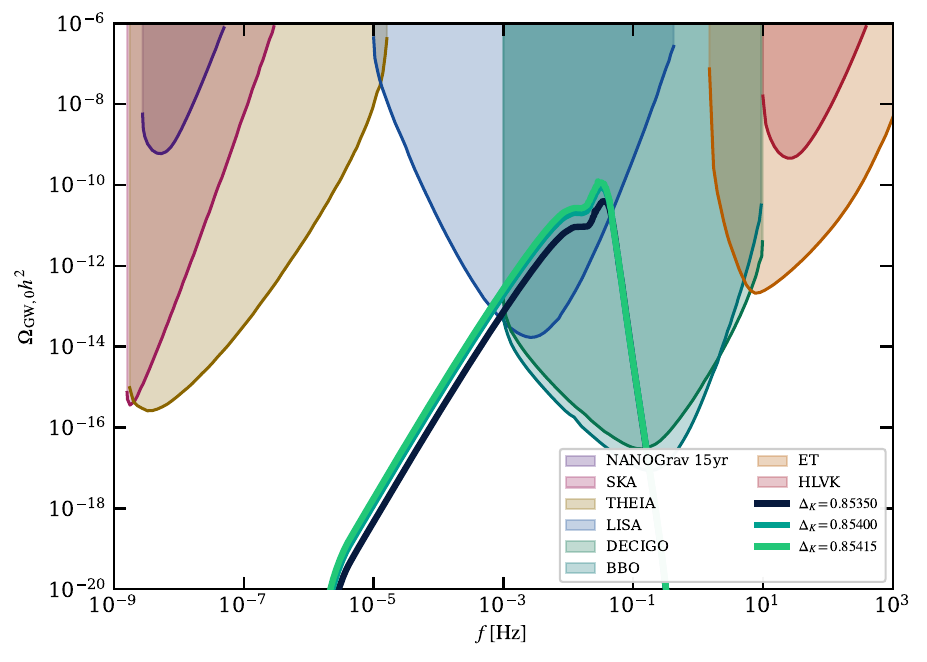}
\caption{Present-day scalar-induced GW spectra generated by the three scalar spectra in \Fig{fig:scalar}. The shaded regions indicate representative experimental sensitivity or frequency coverage~\cite{NANOGrav2023,SKA2015,Theia2021,LISA2017,KawamuraDECIGO2021,BBO2005,EinsteinTelescope2010,HLVK2020}. The feature depth changes the amplitude more strongly than the peak frequency.}
\label{fig:gw}
\end{figure}

Fig\ref{fig:gw} compares the three spectra with representative sensitivity regions from pulsar timing, astrometry, space interferometry, and ground based interferometry~\cite{NANOGrav2023,SKA2015,Theia2021,LISA2017,KawamuraDECIGO2021,BBO2005,EinsteinTelescope2010,HLVK2020}. The nearby feature amplitudes move the GW normalization more strongly than the peak frequency because $u_0$ sets the encounter scale while $\Delta_K$ mainly changes the residence time. The sensitivity curves provide scale comparisons rather than noise-weighted detection forecasts.

\section{Primordial black hole mass function}\label{sec:pbh}
For modes reentering during radiation domination, the smoothed linear density variance is~\cite{YoungByrnesSasaki2014,YoungMusso2020}
\begin{equation}
\sigma_R^2=\frac{16}{81}\int\dd\ln q\,(qR)^4W^2(qR)T^2(qR)\Pz(q),
\label{eq:variance}
\end{equation}
with a real-space top-hat window
\begin{equation}
W(x)=3\frac{\sin x-x\cos x}{x^3}
\end{equation}
and radiation transfer function
\begin{equation}
T(x)=3\frac{\sin(x/\sqrt3)-(x/\sqrt3)\cos(x/\sqrt3)}{(x/\sqrt3)^3}.
\end{equation}

Including the nonlinear curvature-to-density mapping used in compaction based treatments~\cite{YoungMusso2020,DeLucaEtAl2019},
\begin{equation}
\delta=\delta_\ell-\frac38\delta_\ell^2,
\label{eq:nonlinear}
\end{equation}
and take $\delta_c=0.41$ as the reference threshold.Near threshold, critical collapse gives~\cite{Choptuik1993,NiemeyerJedamzik1998,KoikeHaraAdachi1995}
\begin{equation}
M_{\rm PBH}=K M_H(\delta-\delta_c)^{\gamma_c},
\qquad K=3.3,\quad\gamma_c=0.36,
\label{eq:crit}
\end{equation}
with
\begin{equation}
M_H\simeq17\left(\frac{g_*}{10.75}\right)^{-1/6}
\left(\frac{k}{10^6\,{\rm Mpc}^{-1}}\right)^{-2}M_\odot.
\label{eq:MH}
\end{equation}

Instead of attributing the horizon mass to each PBH, we assign the linear density variable to the critical collapse mass and keep the Jacobian. For Gaussian $\delta_\ell$, the mass-weighted formation distribution at a fixed horizon scale can be written
\begin{equation}
\frac{\dd\beta}{\dd\ln M_{\rm PBH}}=
\frac{K}{\gamma_c}\frac{(\delta-\delta_c)^{\gamma_c+1}}
{1-3\delta_\ell/4}
\frac{e^{-\delta_\ell^2/(2\sigma_R^2)}}{\sqrt{2\pi\sigma_R^2}},
\label{eq:dbeta}
\end{equation}
where $\delta$ is understood to be given by \Eq{eq:nonlinear}. Redshifting to the present yields the differential fraction
\begin{equation}
\begin{aligned}
\frac{\dd f_{\rm PBH}}{\dd\ln M_{\rm PBH}}\simeq{}&
1.7\times10^8\left(\frac{g_*}{10.75}\right)^{-1/4}\\
&\times\left(\frac{M_H}{M_\odot}\right)^{-1/2}
\frac{\dd\beta}{\dd\ln M_{\rm PBH}}.
\end{aligned}
\label{eq:fpbh}
\end{equation}

The distribution shown employs the horizon scale in order to maximise the smoothed variance; since critical collapse distributes the masses of PBHs at the time of their formation, the growth factor during the radiation era is a function of $M_H$ and not of each individual final value of $M_{\rm PBH}$.

For the reference spectrum, the Gaussian distribution peaks near
\begin{equation}
\begin{aligned}
M_{\rm PBH}^{\rm mode}&\simeq1.11\times10^{-13}M_\odot,\\
\int\dd\ln M\,\frac{\dd f_{\rm PBH}}{\dd\ln M}&\simeq9.08\times10^{-3}.
\end{aligned}
\label{eq:fint}
\end{equation}

Fig\cite{fig:pbh} compares the Gaussian critical collapse distribution with representative evaporation, lensing, dynamical, accretion, and compact binary limits~\cite{Evaporation2024,Niikura2019,ZumalacarreguiSeljak2018,Brandt2016,Agius2024,LVKSubsolar2022,CarrKohriSendoudaYokoyama2021,GreenKavanagh2021,BaguiEtAl2025}. The maximum lies in the asteroid mass interval. The absolute normalization remains much more sensitive than the peak mass because finite nonattractor statistics and collapse profile dependence modify the rare probability tail~\cite{Maldacena2003,CaiEtAl2018,PassagliaHuMotohashi2019,DeLucaEtAl2019,Caravano2025,Caravano2025II}.
\begin{figure}[t]
\centering
\includegraphics[width=0.98\columnwidth]{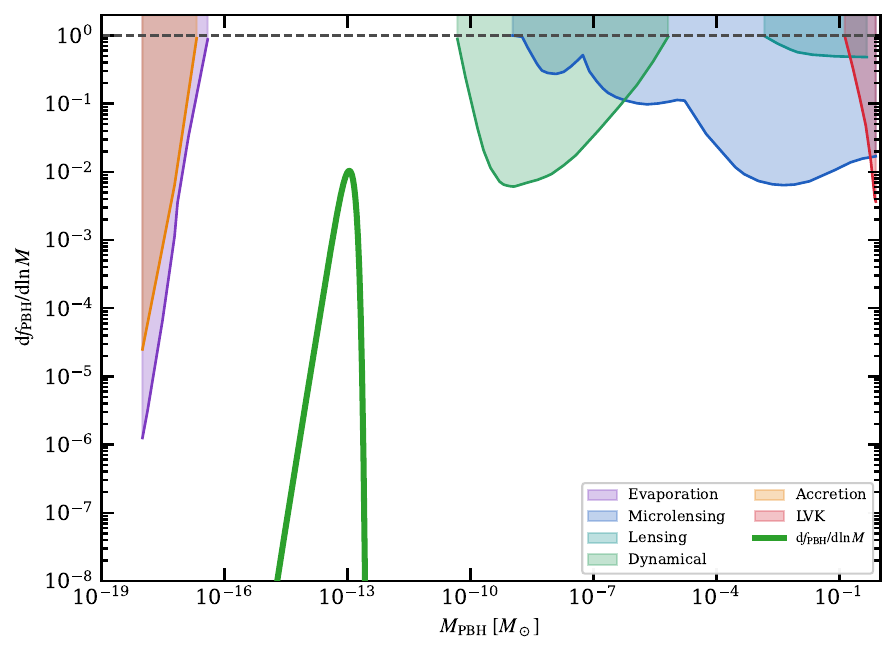}
\caption{Critical collapse PBH mass distribution for the reference numerical scalar spectrum. The green curve is the Gaussian result $\dd f_{\rm PBH}/\dd\ln M$. The comparison regions summarize evaporation~\cite{Evaporation2024}, microlensing~\cite{Niikura2019}, lensing~\cite{ZumalacarreguiSeljak2018}, dynamical~\cite{Brandt2016}, accretion~\cite{Agius2024}, and LVK subsolar mass binary~\cite{LVKSubsolar2022} constraints.}
\label{fig:pbh}
\end{figure}

\section{Discussion and comparison with supersymmetric models}\label{sec:comparison}

The relevant amplification mechanisms are clearly distinguished in the nearby supersymmetric literature. Local Kähler geometry is able to produce a near-inflection region in hybrid or no-scale supergravity~\cite{SpanosStamou2021,WuGongLi2021}. Curved multifield trajectories can lead to an enhancement of the power by means of a different dynamical mechanism~\cite{KawaiKim2023}. Waterfall models can connect the amplified scale to particle masses and to reheating~\cite{AfzalGhoshalKing2025,Khan:2023snv}. GUT Higgs realizations provide the most direct structural comparison since a symmetry-breaking field also carries out the inflationary dynamics~\cite{IjazRehman2025,IjazMehmoodRehman2025,Ijaz:2026ear,MajiMoursyShafi2025}. The present construction combines the radial K\"ahler mechanism with a field direction already assigned a particle physics role in the supersymmetric $331$ model~\cite{KhanLi2026}. The resulting observables organize into a useful hierarchy. CMB quantities depend mainly on the broad nonminimal plateau. The short scale scalar maximum responds strongly to residence time in the localized region. The induced GW signal follows the amplified two-point spectrum with a smoother parameter dependence. The Gaussian PBH abundance responds most sharply because the collapse probability samples the far tail. \Fig{fig:scalar}, \Fig{fig:gw}, and \Fig{fig:pbh} display the hierarchy across the same nearby feature amplitudes. The result shown in \Eq{eq:saddlescale} provides an analytical explanation of the numerical sensitivity; by increasing the amplitude of the features close to the boundary being crossed, the residence time is greatly extended, which accounts for the significantly stronger PBH response as compared to that of the CMB branch. In the effective description, the locally localized K\"ahler operator is kept as an input to the low-energy theory, and at the same time the calculation of neutral stability checks the deformed trajectory within the fields actually involved in inflation.

\section{Conclusions}\label{sec:conclusion}

We studied a localized amplification along the original $331$ breaking inflationary direction of the supersymmetric $SU(6)$ construction. The gauge-invariant K\"ahler deformation lowers the radial force for a finite interval while preserving the parent plateau away from the feature. The resulting slowdown produces a transient nonattractor stage and enhances curvature modes over a narrow range of scales. Dense neutral scans retain positive transverse covariant Hessian eigenvalues throughout the reference trajectory. At the reference pivot, direct mode evolution yields $n_s=0.96647$ and $r=3.25\times10^{-3}$; the scalar maximum attains a value of $5.38\times10^{-3}$ at a wavelength of $1.88\times10^{13}\,{\rm Mpc}^{-1}$. The scalar-induced gravitational wave spectrum associated with this has its peak at $3.17\times10^{-2}\,{\rm Hz}$ and $\Omega_{\rm GW,0}h^2\simeq9.41\times10^{-11}$. The Gaussian critical collapse reference results in an asteroid mass distribution whose integrated fraction is approximately $9.08\times10^{-3}$. The key physical insight is the fact that the various observables differ in their sensitivity to the local geometry. In the case of the CMB branch, the broad plateau continues to prevail. The position of the scalar maximum follows the length of the slowdown period. The GW signal changes smoothly as one passes from one traversing solution to another, while the fraction of PBHs inherits the exponential sensitivity from the rare collapse tail. The critical scaling mentioned in \Eq{eq:saddlescale} accounts for the quick dynamical response observed near the stationary reference pair. The particle sector, the inflationary geometry, and the small-scale observables all constitute one calculation linked to the original $331$ breaking direction.

\begin{acknowledgments}
IK acknowledge financial support from the NSFC under Grant No. 12674335. The authors acknowledge the use of ChatGPT for assistance in improving the English language and clarity of the manuscript.
\end{acknowledgments}

\appendix
\input{appendices}
\bibliographystyle{apsrev4-2}
\bibliography{references}
\end{document}

%% file: appendices.tex
\section{Localized K\"ahler derivatives}\label{app:derivatives}
Writing
\begin{equation}
x=\frac{y^2/2-u_0}{\sigma_u},\qquad s=\operatorname{sech}^2x,\qquad t=\tanh x,
\end{equation}
the radial frame function in \Eq{eq:F} obeys
\begin{align}
F_{,y}&=2\xi y-\frac{2A}{\sigma_u}sty,\\
F_{,yy}&=2\xi-\frac{2A}{\sigma_u}\left[st+\frac{y^2}{\sigma_u}s(1-3t^2)\right].
\end{align}
The radial metric derivative becomes
\begin{equation}
G_{,y}=-\frac{F_{,y}}{F^2}+3\frac{F_{,y}}F\left[\frac{F_{,yy}}F-\left(\frac{F_{,y}}F\right)^2\right].
\end{equation}
The holomorphic combination $f(B)+f(B)^\dagger$ gives
\begin{equation}
\partial_I\partial_{\bar J}\left[f(B)+f(B)^\dagger\right]=0,
\end{equation}
so the localized term changes the radial force through $F$ and its first derivative without introducing an additional mixed K\"ahler derivative on the real trajectory. The phase stabilizer vanishes at $S=0$ and $B=B^\dagger$, leaving the radial pullback unchanged.

\section{Neutral covariant Hessian and PBH Jacobian}\label{app:hessian}
Writing the three complex fields in six real coordinates gives a real kinetic matrix $G_{AB}$. The covariant potential Hessian is
\begin{equation}
\nabla_A\nabla_BV=\partial_A\partial_BV-\Gamma^C{}_{AB}\partial_CV.
\end{equation}
The generalized eigenvalue problem
\begin{equation}
\det\!\left(\nabla_A\nabla_BV-m^2G_{AB}\right)=0
\end{equation}
provides the transverse potential masses shown in \Fig{fig:stability}. The tangent direction and broken gauge orbit are projected out with the $G_{AB}$ inner product before taking the lightest physical transverse eigenvalue.

For critical collapse, writing $x=\delta(\delta_\ell)-\delta_c$ gives
\begin{equation}
M_{\rm PBH}=K M_Hx^{\gamma_c},
\qquad
\frac{\dd\ln M_{\rm PBH}}{\dd\delta_\ell}
=\gamma_c\frac{1-3\delta_\ell/4}{x}.
\end{equation}
The mass weighted formation probability contains $M_{\rm PBH}/M_H=Kx^{\gamma_c}$. Multiplication by the inverse Jacobian yields \Eq{eq:dbeta}. The radiation era growth from formation to equality depends on the formation horizon mass $M_H$, while critical collapse distributes the resulting PBHs over $M_{\rm PBH}$ at the same formation epoch.